\documentclass[12pt,a4paper,titlepage,leqno]{article}
\usepackage[T1]{fontenc}
\usepackage{graphicx}
\usepackage{mathtools}
\usepackage{amssymb}
\usepackage{amsthm}
\usepackage{xcolor}
\usepackage{hyperref}
\usepackage{natbib}
\usepackage{xspace}

\newcommand{\ie}{{\rm i.e.}\xspace}

\newcommand{\Tabref}[1]{\mbox{Table~\ref{#1}}}

\newcommand{\tabref}[1]{\mbox{Table~\ref{#1}}}

\newcommand{\scepg}{\texttt{scEPG}\xspace}
\newcommand{\scepgs}{\texttt{scEPG}s\xspace}

\title{Searching for related contributors to a DNA mixture using single cell analysis}
\author{Robert G.  Cowell}

\begin{document}
	\maketitle
	\begin{abstract}
		We propose a staged process for assessing the presence of related individuals in an autosomal STR DNA mixture sample, when it is analysed using a single cell pipeline analysis (as opposed to a bulk mixture analysis). In the absence of a large dataset if single cells from real related people, artificial single cell data is generated by sampling individual locus data taken from single cell dataset of real unrelated people. It is applied to create random admixtures used in a relatedness search for pairs of individuals to illustrate the usefulness of the method.
	\end{abstract}

\section{Introduction}	

There have been large strides in the theoretical analysis of STR DNA mixtures in the last two decades, in the development and use of software, which implements  the theoretical probabilistic methods, is now well accepted and established by courts in multiple jurisdictions around the world. Much of the work in the  implementation of such \textit{probabilistic genotyping} software assumes that contributors to a mixture sample are assumed to be unrelated; however in recent years some progress has been made in dropping this restriction, such as
\cite{Green2020KinMix,GreenMortera2021,Hernandis2019Relmix,Dorum2017,Taylor2013,Taylor2014,bleka2025efmex}.

Each of these software systems treat mixture samples as a bulk item, in which multiple cells within a mixture are sampled and analysed together. Although well developed, they have problems if the mixtures have many people contributing DNA, with five or more contributors making mixtures  computationally expensive to analyse; they also depend on the amounts of DNA from each assumed untyped contributor to be quite different in order to separate and identify their genetic profiles.

In the last few years an alternative approach to analysing STR DNA mixtures  has been proposed: single cell genetic analysis. A very recent review is \cite{ballantyne2026next}. In this approach individual cells are extracted from the mixture, and amplified in isolation. This results in a set of single cell electropherograms (\scepgs), numbering several dozen or more. Each cell comes from a single contributor, and the job is to group these \scepgs into clusters, using the peak height information in the \scepgs, so that all cells in each cluster come from one person, and the cells in distinct clusters come from different people. Approaches to do this have been proposed using probabilistic models, see \citep{duffy2023evidentiary,qhawe2025,cowell2025clustering}, which are very efficient and accurate provided that the \scepgs do not have a lot of dropout or other artefacts. However, these clustering methods are based on the contributors whose DNA is amplified in the selected single cells are (assumed to be) unrelated.

In this paper we combine the method of \cite{cowell2025clustering}, for cell clustering of (assumed) unrelated contributors to a mixture using \scepgs, with the proposed pedigree reconstruction method of \cite{cowell2003clustering}. The combination is intended to be used to search for  relatedness between contributors to a set of  \scepgs,  and also possibly of  relating non-contributors to contributors.

 The plan of  the paper is as follows. The next section summarizes the clustering approach of \cite{cowell2003clustering} for pedigree reconstruction and the
partition search of \cite{cowell2025clustering} for clusters in set of \scepgs, and how to combine the algorithms to search for related contributors to a set of \scepgs. We then discuss a dataset of \scepgs\ of presumed unrelated individuals, and how to use the dataset simulate pairs of closely related individuals and \scepgs from them. We then report on the results of simulations involving 2000 simulated admixtures using the simulated \scepgs\ and the known genotypes of the simulated contributors to these \scepgs.

\section{Theoretical overview}

\subsection{IBD clustering}
The references cited in the introduction, that combine probabilistic genotyping mixture analysis with relatedness, assume that the relatedness of individuals  is specified by a particular pedigree, which may include extra persons to explain the relatedness. For example we might introduce parents to relate sibling contributors to a mixture. 

The approach taken in \cite{cowell2003clustering}, for reconstructing pedigrees of typed persons, does not assume a pedigree structure, but attempts to find a small subset of pedigrees that could explain relatedness of individuals. The particular application in mind of the paper was to look for small groups of related individuals among a large group of typed individuals; such an analysis may be of use in a mass disaster scenario, or grouping together family members in a mass grave.

The paper proposed a two stage process. In the first stage, distinct pairs of individuals are taken, and their IBD distance (as defined in the paper: essentially the number of meioses in the--assumed non-inbred\footnote{By non-inbred, we mean that the founders of the pedigree are unrelated, and that there is only one path of meioses connecting the two individuals. Thus for example we exclude children of cousins.}- pedigree connecting the two individuals) is estimated from their genotypes and population allele frequencies. 
In essence, given the  two people,  the probability that they share an allele identical by descent (IBD) for a distance $k$ is evaluated on a specific locus. We may denote this by $p_{m,k}$ where $m$ denotes the locus, and $k$ the distance. The probability that they do not have an allele IBD is $1 - p_{m,k}$. If the two individuals do not have an allele in common in their genotypes on the locus, then we have the probability $1 - p_{m,k}$; however if they have an allele in common, then it is IBS (identical by state) but may or  not be IBD, and the probability is a weighted combination of $1 - p_{m,k}$ and $p_{m,k}$, where the weights depend upon and can be found from the genotypes of the two people. We thus have a probability of the form
$a(m)p_{m,k} + b(m)(1-p_{m,k})$ where the coefficients $a(m)$ and $b(m)$ depend upon the genotypes of the two people on the locus $m$. For a set of unlinked autosomal loci we obtain a likelihood
$$L(k \vert gt_1, gt_2) = \prod_m\left( a(m)p_{m,k} + b(m)(1-p_{m,k})\right)$$
We evaluate this for values $k=1,2,3,\ldots,$ up to some limit say $k=10$, giving a likelihood distribution over $k$, from which we find the $k$ that maximizes the likelihood. The smaller the value of $\hat{k}$, the closer the relationship between the two people is estimated to be, large values indicate unrelatedness. It was from simulations found that, due to high variance of the likelihood distance distribution, only values of $\hat{k}$ of the order of 3 or less proved reasonably reliable; however this is sufficient for our purpose.

So for a collection of people $n$ people we find all distinct pairs of likelihood distributions $L(k \vert gt_i, gt_j)$, and from this find the value $\hat{k}$ that maximizes this likelihood. For the second stage  we construct an \textit{undirected graph} in which the nodes are the people, and we join the people if their estimated distance  $\hat{k} \le 3$.  This undirected graph may be used to construct pedigrees that are compatible with it - typically they will number a much smaller number that the pedigrees that could be constructed from the $n$ people without constraint. For example suppose we have three people, labelled by  1, 2 and 3, and we estimate  pairwise distances  $\widehat{k_{1,3}}= \widehat{k_{2,3}} = 1$ and  $\widehat{k_{1,2}} = 10$. Then this could strongly suggest that persons 1 and 2 are unrelated parents of person 3.  On the other hand if we estimate
 $\widehat{k_{1,3}}= \widehat{k_{2,3}} = 1$ and  $\widehat{k_{1,2}} = 2$, this could indicate that person 3 is a parent to  sibling or half-sibling persons 1 and 2.  It could also indicate that persons 1 and 2 are grandparent-grandchild, and 3 is a child of the grandparent and parent of the child.
  This stage was not carried out in the paper in a general way, but examples, including the Romanov data \citep{gill1994identification} were given.  

In \cite{cowell2003clustering} it was more convenient to evaluate the likelihood ratio of a pair of people having distance $k$ to them being unrelated (ie., having distance $\infty$). Also note that special consideration for the distance definition of two full sibs was required, because full sibs may have 2 alleles that are IBD: in the paper a pair of full-sibs is defined to have distance 1.5. Assuming Hardy-Weinberg equilibrium, and no inbreeding of the individuals in the (unknown) pedigree  relating the individuals of interest,  \cite{cowell2003clustering} show the likelihood ratios are given as shown in Table~\ref{tab:lrat1}.

\begin{table}[htbp]
	\begin{center}
		\caption{The likelihood ratios of different distances (versus infinite distance), for
			different types of data for two (unordered) persons, assuming Hardy-Weinberg equilibrium and no inbreeding. The $p_a$ and $p_b$ are population allele frequencies. \vspace{12pt}}
		\begin{tabular}{|l|c|c|}  \hline
			Genotype cases: & Distance $k\in{\mathbb N}$ & Distance 1.5 \\ \hline
			(1) aa, aa  \phantom{\Large $\int_x^x$}
			& $(\frac12)^{k-1}\frac1{p_a}+1-(\frac12)^{k-1}$ & $\frac14\frac1{p_a^2}+\frac12\frac1{p_a}+\frac14$ \\ \hline
			(2) aa, ab  \phantom{\Large $\int_x^x$}
			& $(\frac12)^{k-1}\frac1{2p_a}+1-(\frac12)^{k-1}$ & $\frac12\frac1{2p_a}+\frac14$ \\ \hline
			(3) ab, ab  \phantom{\Large $\int_x^x$}
			& $(\frac12)^{k-1}\frac{p_a+p_b}{4p_ap_b}+1-(\frac12)^{k-1}$ & $\frac14\frac1{2p_ap_b}+\frac12\frac{p_a+p_b}{4p_ap_b}+\frac14$ \\ \hline
			(4) ab, ac  \phantom{\Large $\int_x^x$}
			& $(\frac12)^{k-1}\frac1{4p_a}+1-(\frac12)^{k-1}$ & $\frac12\frac1{4p_a}+\frac14$ \\ \hline
			(5) no alleles in common \phantom{\Large $\int_x^x$}
			& $1 - (\frac12)^{k-1}$ & $\frac14$   \\ \hline
		\end{tabular}
		\label{tab:lrat1}
	\end{center}
\end{table}

\subsection{{\scepg} clustering}

\cite{cowell2025clustering} proposed a probabilistic clustering algorithm  for the analysis of
forensic DNA mixtures in which individual cells are isolated, and short
tandem repeats are amplified using the polymerase chain reaction to
generate single cell electropherograms ({\scepgs}). The task of the algorithm was 
to use the peak height information in the electropherograms to group
the cells according to their contributors. The proposed algorithms work whether or not there are typed persons of interest who are suspected of being in the collection of \scepgs. The result of running the algorithm is a set of partitions of the \scepgs, ordered by  their decreasing likelihoods of the observed peaks heights.
Using a recently developed
experimental set of individual cell electropherograms, the paper presented the results of a large set of
simulations that showed the proposed clustering algorithm has excellent performance in correctly grouping single cells, and for assigning
likelihood ratios for types persons of interest.

Typically the difference in ${\log_{10}}$-likelihoods between the highest likelihood partition and the next partition is around 10, so most of the time we may take that partition as the estimate of the true partition in which: (a) all cells within a cluster of the partition come from the same person, and (b) cells in distinct clusters come from different people. 
In some cases, there the number of cells from a person is small and has large amount of dropout, there may be several partitions whose likelihoods are close - in that case   averaging may be applied over the partitions using their normalized likelihoods as posterior probabilities of being the correct partition (assuming at that one of them is). In many simulations (tens of thousands) the highest likelihood partition found was the true partition.

However this was done so under the assumption that the contributors are unrelated. We now attempt to remedy this. For brevity, we shall only deal with using the highest likelihood partition in the following explanation, but the simulation results later will make use of a set of high likelihood partitions.

When the highest likelihood partition of the \scepgs\ is found, other  information besides the likelihood value is available from the computational process. To begin with, the number of clusters in the partition is the estimated number of contributors. For each cluster a log-likelihood is found - the partition log-likelihood is the sum of those of its partitions. That is, we treat the \scepgs of the cells as independent given their contributors. There is a slight complication here, discussed in \cite{cowell2025clustering}, in dealing with population sub-structure corrections. Strictly we cannot evaluate the cluster log-likelihoods separately and add them together if we \textit{fully} take account of substructure corrections. We follow \cite{cowell2025clustering} in applying sub-structure  correction within a cluster, but not across clusters. This, it is argued, is an approximation with a small relative error and facilitates computational evaluations.

Now within each cluster, and each locus, we need to compute for each cell the likelihood $p(h_{i,m} | gt_m)$  for each genotype $gt_m$ and peak height information $h_{i,m}$ for locus $m$ on cell $c_i$ in the cluster. The total likelihood for the cluster is given by

$$\prod_m \left(\sum_{gt_{i,m}} p(gt_{i,m})\prod_i p(h_{i,m} \vert gt_{i,m}) \right).$$

If we normalize to unity the terms in the sum for each locus, we will obtain the posterior probability for the contributor genotype on the  locus,

\begin{equation}
 p(gt_{i,m}\vert \{h_{i,m}\}) \propto p(gt_{i,m})\prod_i p(h_{i,m} \vert gt_{i,m}),
\label{eq:gtpost}
\end{equation}
where the product is over all cells in cluster $i$ (the $i$ suffix on the gt is not needed for this, but we add it for when we take a pair of clusters),
in which population sub-structure correction is applied to the genotype probabilities $p(gt_{i,m})$. The priors and posteriors are thus readily computed for each locus.

For a distinct pair of clusters $c_i$ and $c_j$ we will have on each locus the joint posterior distribution given by the product
$$ p(gt_{i,m}\vert \{h_{i,m}\})  p(gt_{j,m}\vert \{h_{j,m}\})$$
because we have assumed the contributors are unrelated.

\subsection{Combining the algorithms}

We may now combine the algorithms. Suppose we have a set of \scepgs, and that we do not have any typed persons of interest. We first find the highest likelihood partition, assuming contributors are unrelated. Then, for each cluster in  the partition, we compute for each locus the posterior distribution for the contributor's genotype. 

Suppose also  we had assumed Hardy-Weinberg equilibrium for the partition search. Then for each pair of genotypes $gt_{i,m}$ and $gt_{j,m}$ we look up in Table~\ref{tab:lrat1} the appropriate matching genotype case, and thus the likelihood ratio given distance $k$. Let us denote this by $LR(k\vert gt_{i,m}, gt_{j,m})$. We   form the weighted sum

\begin{equation}
\sum_{gt_{i,m}} \sum_{gt_{j,m}} 
p(gt_{i,m}\vert \{h_{i,m}\})  p(gt_{j,m}\vert \{h_{j,m}\})
LR(k\vert gt_{i,m}, gt_{j,m})
\label{eq:lrcor}
\end{equation}

This is found for each locus, and their products formed. 
But now recall that for each cluster and locus,
$$ p(gt_{im}\vert \{h_{i,m}\}) \propto p(gt_{i,m})\prod_i p(h_{i,m} \vert gt_m).$$
In addition, in finding the entries in Table~\ref{tab:lrat1}, the denominator was simply the joint genotype probability under the assumptions of the two people being unrelated, that is, the joint genotype $p(gt_{i,m})p(gt_{j,m})$. Hence, essentially, this denominator cancels with its appearance in the numerator of (\ref{eq:lrcor}), so that the sum is a correction factor to the combined likelihoods of the clusters given the distance $k$. Evaluating this for each $k$, multiplying the values over each locus, we can thus find the value of $k$ which maximizes this correction factor.

If we take account of population sub-structure in the partition search, the calculations and arguments go through the same, except that the entries in Table~\ref{tab:lrat1} need to be modified to take account of population-substructure as well. This is relatively straightforward, a summary of relevant formulae is given in Appendix~\ref{sec:app1}; we mention that in the simulations that are described later sub-structure was included.

\subsection{Comments}

If the IBD distance estimates find only two cluster contributors in the partition may be related, there is no need to consider constructing a pedigree, unless other non DNA information is available, for example, was it expected there is a mother and son in the mixture, and it is required to find the likelihood ratio for this pedigree (versus son not in the mixture).

The above has been framed in terms of having no typed persons in the partition search. The partition search allows for some people to be typed - for such people the prior genotype probability will have a single non-zero term 
$p(gt_{i,m})$ thus simplify the calculations.

We may also be interested in the case of a typed person who is not in the mixture, but may have a relative in the mixture. The evaluations proceed similarly, but now  the correction factor is simpler as there is no summation over the typed person's genotypes ($j$ is the untyped person in the mixture):

\begin{equation}
	 	p(gt_{i,m} ) \sum_{gt_{j,m}}  p(gt_{j,m}\vert \{h_{j,m}\})
	LR(k\vert gt_{i,m}, gt_{j,m})
	\label{eq:lrcor2}
\end{equation}

If there are several pairs of close relationships found, one can proceed to construct the undirected graph, and thence various possible pedigrees consistent with it and possibly DNA unrelated information. One can then combine the joint -genotype pedigree likelihood with the cluster likelihoods $\prod_i p(h_{i,m} \vert gt_m)$ with each such cluster likelihood identified with at particular person in the pedigree via the undirected graph. Bayesian network formulations simplify this meshing of the two computations. We do not pursue this further in this paper -- simply due to lack of data -- which brings us to the next section, on the data to use in the simulations.

\section{Data on relatives}

Before discussing the simulations, we first need to discuss the data used in the simulations. 
We use the population data for Caucasians taken from \cite{butler:etal:03} for all allele frequencies. For the \scepgs, we use the same set of data from \cite{duffy2023evidentiary}, also used in  \cite{cowell2025clustering}, consisting of nearly 3000 \scepgs 
split into two groups. A
training set of nearly 2000 \scepgs generated in various numbers from samples of people having 41 distinct genotypes was used to calibrate
the peak height probability model. The remaining formed the test set of around 1000 \scepgs from 13 individuals used to simulate the genotypes, \scepgs, and admixtures of related individuals.  As the genotypes are known, this was tested by estimating  all the pairwise IBD distances between the people, using distance values up to 10, both with and without population sub-structure. With 54 people there are 1431 possible pairings, so one might expect by chance to be some low distance value estimates. To gauge how many to expect, and compare to the number observed, 100 samples of genotypes of 54 people were sampled using the population, and the average number of estimated distances were found, the results are shown in \tabref{tab:maxibd}. The real and simulated  people values are similar both without and with sub-structure correction, with the possible exceptions of $\hat{k}=4$ for $\theta=0$, and $\hat{k}=5$ for $\theta = 0.01$. However, as discussed earlier  values of $\hat{k}\ge 4$ are quite consistent with people being unrelated.

\begin{table}[htbp]
	\caption{Pairwise maximum IBD distance counts for genotypes of  people in scEPG dataset (columns 2 and 5) and averages from 100 simulations of 54 random people (columns 3 and 6). }
	\begin{center}
		\begin{tabular}{c|rr|rr}\hline
			Distance &\multicolumn{2}{c|}{Real People $(k)$}&
			\multicolumn{2}{c}{Simulated People $(\hat{k})$}\\
			$k$ or $\hat{k}$& $\theta$=0 &  $\theta$=0.01 & $\theta$=0 &  $\theta$=0.01  \\ \hline
1	&	0	&	0	&	0.01	&	0.00	\\
1.5	&	0	&	0	&	0.30	&	0.22	\\
2	&	32	&	23	&	29.03	&	20.46	\\
3	&	158	&	119	&	159.87	&	118.70	\\
4	&	227	&	161	&	185.65	&	149.11	\\
5	&	120	&	139	&	122.58	&	104.02	\\
6	&	54	&	58	&	65.07	&	60.37	\\
7	&	27	&	20	&	32.56	&	31.21	\\
8	&	23	&	15	&	16.37	&	15.91	\\
9	&	10	&	11	&	7.60	&	7.40	\\
10	&	780	&	885	&	811.96	&	923.60	\\
			\hline
		\end{tabular}
	\end{center}
	\label{tab:maxibd}
\end{table}

So the question arises: How to do a simulation study for the combined partitions search and IBD-distance estimation procedure for related individuals in the absence of a large dataset of related individuals?

The answer we use is that from the testing dataset of real \scepgs from the real people, we simulate genotypes of people that are related in some manner, and then for each simulated person simulate an \scepg by sampling without replacement the  peak height data from a real \scepg on that locus from a real person matching the genotype of the simulated person on  that locus. In essence, we are treating the individual locus \scepg values as interchangeable. This seems reasonable, because in evaluating peak height likelihoods in carrying out the partition search we assume (in our likelihood model for the peak heights) that the \scepgs of separate cells are independent given the genotypes of the contributors, and that when evaluating the peak height likelihoods for an individual cell that the heights from different locus are independent.

Another  issue here is that we cannot simply simulate random genotypes for people from the general population, because we require that for each simulated genotype we can find a matching real person so that we can use one of their \scepg\ data on that locus when constructing a random \scepg\ for the random person. Hence we must sample from the set of genotypes on each locus of the individuals in the real \scepg\ dataset to ensure we may construct valid simulated \scepgs. However, in drawing from a much smaller pool of genotypes we may introduce some bias in the simulations.

The simulation process had two stages: first generate possible genotypes; and then for the possible genotypes generate the  artificial \scepgs.

With this, we simulated genotypes of pairs of people for the following types of relationships, with mutation excluded:

\begin{itemize}
	\item Parent-child;
	\item Grandparent-grandchild;
	\item Half-siblings;
	\item Full-siblings.
\end{itemize}

Also,  Mother-Father-Child were trios were generated. 
 For each type of relationship 100 different genotypes were simulated, and for each such simulated set of genotypes the maximum number of \scepgs\ were generated that could be sampled without replacement from the set of test \scepgs.
 
 Simulating a parent-child genotype pair (distance 1) is straightforward, because they have exactly one allele IBD in common. So for each locus, we first sample a random genotype from the real people, and select one of the two alleles to be IBD with the child, say $A$. We then sample uniformly from all the other people who have at least one $A$ in their genotype. We do this for each locus. (Note that although this gives alleles that are IBS, but does not imply that they are IBD.)
 
 Sampling of grandparent-grandchild and half-siblings is similar, because they have both distance 2. This means that they have probability 0.5 of having 0 alleles IBD, and probability 0.5 that they have one allele IBD. So on a locus we sample a genotype from the real people. We then do a coin toss to decide if the other person will have an allele IBD; if no then we sample any other person's genotype on the locus, if yes then we sample proceed as the parent-child: sample an allele $A$ to be IBD, and then sample from all the other genotypes that have at least one allele of type $A$ in their genotype.
 
 Full siblings are slightly more complex. There is a probability 0.25 that two siblings have no alleles IBD, probability 0.5 that they have one allele IBD, and probability 0.25 that they have two alleles IBD (assuming no inbreeding). So we proceed as follows. For a locus, sample the genotype for one sibling. Then sample a value $p$ uniformly from $[0,1]$. If $p< 0.25$ then sample any other person's genotype; else if $p < 0.75$ sample any other person's genotype that shares at least one allele of the same type as the sibling; otherwise ($p \ge 0.75$) sample any other person's genotype that is the same as the first sibling has.
 
 Simulating a trio proceeds as follows. 
 For each locus, sample the child's genotype from the genotypes of the real persons. Then for each locus allele in the child's genotype sample the genotype of a parent such that the parent has at least one copy of the allele.

 \subsection{Testing the simulated genotypes}
 
 Before looking at the clustering using the simulated \scepgs, we first examining the quality of the simulated genotypes. This is  important because they are not being simulated from the full population distribution of alleles, but from the subset of  alleles in the genotypes of the contributors to the real \scepgs.
 
 \subsubsection{Two-person relatedness}
 
 For each of the two person simulated profiles, 100 in each type, the simulated genotypes were used to find the maximized likelihood IBD distance. The results are shown in \Tabref{tab:prchk}, with values at the true distance of each underlined. (No subs-structure correction : $\theta=0$.)
 
 \begin{table}[htbp]
 	\caption{Counts of estimated distances for each of the four closely related pairs of simulated individuals, for 100 simulated pairs of people for each type. The underlined counts indicate the true distances in each case.}
 	\begin{center}
 		\begin{tabular}{c|cccc}\hline
 			Distance & Full Sibs & Parent-Child & Half Sibs & Grandparent-grandchild  \\ \hline
 			1 & 1 & \underline{89} & 4 & 3  \\
 			1.5 & \underline{94} & 11 & 25 & 19  \\
 			2 & 5 & 0 & \underline{56} & \underline{51}  \\
 			3 & 0 & 0 & 11 & 17  \\
 			4 & 0 & 0 & 3 & 6  \\
 			5 & 0 & 0 & 0 & 3  \\
 			6 & 0 & 0 & 0 & 0  \\
 			7 & 0 & 0 & 0 & 1  \\
 			8 & 0 & 0 & 0 & 0  \\
 			9 & 0 & 0 & 0 & 0  \\
 			10 & 0 & 0 & 1 & 0  \\
 			\hline
 		\end{tabular}
 	\end{center}
 	\label{tab:prchk}
 \end{table}
 
 We see that on the whole the correct distance is found in most cases, and almost all within a distance $\pm 1$.

 \subsubsection{Trio relatedness}
 
 As the ``gold standard'' we  simulate family trios from the full genotype distribution. This is then compared to genotypes simulated from the full dataset, and from the set of test persons. For each of the three sets of simulations, in each trio the maximum distance is recorded, and the normalized distance distribution is found. This gives a maximized distance distribution for each simulation, and from the set normalized distributions and overall normalized distribution may be found. 
 
 For simplicity of carrying out the simulation, the evaluations assumed Hardy-Weinberg equilibrium of population alleles (\ie, $\theta = 0$).
 
 The results for the full distribution are show in \tabref{tab:ppctriofull}, based on 1000 simulated trios. We see that both father-child and mother-child parental relationships are identified correctly with distance 1, with a small number in each mis-identified as being of siblings. If we take a distance of 5 or more as indicating effectively unrelated, we see that most of the father-mother distances are correctly identified as unrelated, as they were simulated to be. However we still see some mis-classifications, and the table therefore indicates the frequency of these to be expected by chance.  
 
 \begin{table}[htbp]
 	\caption{Distributions of maximized distance and averaged distance distributions for each pair of individuals in a father-mother-child trio. Trios generated from the full genotype distribution on each locus.}
 	\begin{center}
 		\begin{tabular}{c|cc|cc|cc}\hline
 			&\multicolumn{2}{c|}{Father Mother}&
 			\multicolumn{2}{c|}{Father Child}&
 			\multicolumn{2}{c}{Mother Child}\\
 			Distance $i$& Max&Average& Max&Average& Max&Average\\ \hline
 			1 & 0 & 0 & \underline{984} & 0.958114 & \underline{989} & 0.961599  \\
 			1.5 & 0 & 0.000526028 & 16 & 0.0375072 & 11 & 0.0345702  \\
 			2 & 23 & 0.017213 & 0 & 0.00420085 & 0 & 0.00370985  \\
 			3 & 104 & 0.0389186 & 0 & 0.000129916 & 0 & 9.58E-05  \\
 			4 & 142 & 0.0491151 & 0 & 1.66E-05 & 0 & 1.04E-05  \\
 			5 & 106 & 0.0531498 & 0 & 5.35E-06 & 0 & 2.93E-06  \\
 			6 & 44 & 0.0548202 & 0 & 2.94E-06 & 0 & 1.48E-06  \\
 			7 & 24 & 0.0555675 & 0 & 2.16E-06 & 0 & 1.04E-06  \\
 			8 & 9 & 0.0559199 & 0 & 1.85E-06 & 0 & 8.62E-07  \\
 			9 & 6 & 0.056091 & 0 & 1.71E-06 & 0 & 7.86E-07  \\
 			10 & 2 & 0.0561752 & 0 & 1.65E-06 & 0 & 7.50E-07  \\
 			11 & 1 & 0.0562171 & 0 & 1.62E-06 & 0 & 7.33E-07  \\
 			12 & 0 & 0.0562379 & 0 & 1.60E-06 & 0 & 7.24E-07  \\
 			13 & 1 & 0.0562483 & 0 & 1.59E-06 & 0 & 7.20E-07  \\
 			14 & 0 & 0.0562535 & 0 & 1.59E-06 & 0 & 7.18E-07  \\
 			15 & 0 & 0.0562561 & 0 & 1.59E-06 & 0 & 7.17E-07  \\
 			16 & 0 & 0.0562574 & 0 & 1.58E-06 & 0 & 7.16E-07  \\
 			17 & 0 & 0.056258 & 0 & 1.58E-06 & 0 & 7.16E-07  \\
 			18 & 0 & 0.0562584 & 0 & 1.58E-06 & 0 & 7.16E-07  \\
 			19 & 0 & 0.0562585 & 0 & 1.58E-06 & 0 & 7.16E-07  \\
 			20 & \underline{538} & 0.0562586 & 0 & 1.58E-06 & 0 & 7.16E-07  \\
 			\hline
 		\end{tabular}
 	\end{center}
 	\label{tab:ppctriofull}
 \end{table}
 
 For the trios randomly generated using all the test and training people, 
 the results shown in \tabref{tab:ppctriotestandtrain} again show mostly correct parental relatedness in most cases, but now there is an increase in mis-classifications as full siblings. 
 While the parent-child distances are correctly found most of the time, the two parent, who should be unrelated, exhibit a large number of appearing to be closely related.
  \begin{table}[htbp]
 	\caption{Distributions of maximized distance and averaged distance distributions for each pair of individuals in a father-mother-child trio. Trios generated from the genotypes occurring in both the test and training datasets.}
 	\begin{center}
 		\begin{tabular}{c|cc|cc|cc}\hline
 			&\multicolumn{2}{c|}{Father Mother}&
 			\multicolumn{2}{c|}{Father Child}&
 			\multicolumn{2}{c}{Mother Child}\\
 			Distance $i$& Max&Average& Max&Average& Max&Average\\ \hline
 			1 & 0 & 0 & \underline{923} & 0.867852 & \underline{915} & 0.867421  \\
 			1.5 & 4 & 0.0056506 & 77 & 0.124891 & 85 & 0.12543  \\
 			2 & 121 & 0.0545598 & 0 & 0.00678887 & 0 & 0.00654046  \\
 			3 & 231 & 0.0655057 & 0 & 0.000310684 & 0 & 0.000326914  \\
 			4 & 156 & 0.0604074 & 0 & 4.95E-05 & 0 & 6.36E-05  \\
 			5 & 90 & 0.0557751 & 0 & 1.75E-05 & 0 & 2.75E-05  \\
 			6 & 45 & 0.053068 & 0 & 9.98E-06 & 0 & 1.81E-05  \\
 			7 & 22 & 0.051632 & 0 & 7.43E-06 & 0 & 1.47E-05  \\
 			8 & 8 & 0.0508951 & 0 & 6.39E-06 & 0 & 1.32E-05  \\
 			9 & 7 & 0.0505221 & 0 & 5.92E-06 & 0 & 1.25E-05  \\
 			10 & 5 & 0.0503345 & 0 & 5.69E-06 & 0 & 1.22E-05  \\
 			11 & 1 & 0.0502404 & 0 & 5.58E-06 & 0 & 1.21E-05  \\
 			12 & 0 & 0.0501933 & 0 & 5.53E-06 & 0 & 1.20E-05  \\
 			13 & 1 & 0.0501697 & 0 & 5.50E-06 & 0 & 1.19E-05  \\
 			14 & 0 & 0.0501579 & 0 & 5.49E-06 & 0 & 1.19E-05  \\
 			15 & 0 & 0.050152 & 0 & 5.48E-06 & 0 & 1.19E-05  \\
 			16 & 0 & 0.0501491 & 0 & 5.48E-06 & 0 & 1.19E-05  \\
 			17 & 0 & 0.0501476 & 0 & 5.48E-06 & 0 & 1.19E-05  \\
 			18 & 0 & 0.0501469 & 0 & 5.48E-06 & 0 & 1.19E-05  \\
 			19 & 0 & 0.0501465 & 0 & 5.47E-06 & 0 & 1.19E-05  \\
 			20 & \underline{309} & 0.0501463 & 0 & 5.47E-06 & 0 & 1.19E-05  \\
 			\hline
 		\end{tabular}
 	\end{center}
 	\label{tab:ppctriotestandtrain}
 \end{table}
  
 For the trios from the people generated from the test persons only, 
 the values shown in \tabref{tab:ppctriotest} show much worse predictions. Now over 60\% of the parental relationships are mis-identified as being full-siblings, and for the two parents only 1\% of the simulations have a distance of 5 or more, with over a third predicting a full-sib relatedness between the two parents.
 
 \begin{table}[htbp]
 	\caption{Distributions of maximized distance and averaged distance distributions for each pair of individuals in a father-mother-child trio. Trios generated from the genotypes occurring only in the test datasets.}
 	\begin{center}
 		\begin{tabular}{c|cc|cc|cc}\hline
 			&\multicolumn{2}{c|}{Father Mother}&
 			\multicolumn{2}{c|}{Father Child}&
 			\multicolumn{2}{c}{Mother Child}\\
 			Distance $i$& Max&Average& Max&Average& Max&Average\\ \hline
 			1 & 21 & 0.0186922 & \underline{390} & 0.401696 & \underline{377} & 0.389561  \\
 			1.5 & 350 & 0.303733 & 610 & 0.595776 & 623 & 0.608135  \\
 			2 & 459 & 0.238054 & 0 & 0.00240901 & 0 & 0.00220269  \\
 			3 & 139 & 0.0997025 & 0 & 8.66E-05 & 0 & 7.50E-05  \\
 			4 & 21 & 0.0475134 & 0 & 1.16E-05 & 0 & 9.69E-06  \\
 			5 & 5 & 0.0295533 & 0 & 3.67E-06 & 0 & 3.02E-06  \\
 			6 & 1 & 0.0225542 & 0 & 1.96E-06 & 0 & 1.61E-06  \\
 			7 & 0 & 0.019517 & 0 & 1.40E-06 & 0 & 1.15E-06  \\
 			8 & 0 & 0.0181086 & 0 & 1.18E-06 & 0 & 9.72E-07  \\
 			9 & 0 & 0.0174313 & 0 & 1.09E-06 & 0 & 8.91E-07  \\
 			10 & 0 & 0.0170992 & 0 & 1.04E-06 & 0 & 8.53E-07  \\
 			11 & 0 & 0.0169348 & 0 & 1.02E-06 & 0 & 8.35E-07  \\
 			12 & 0 & 0.0168531 & 0 & 1.01E-06 & 0 & 8.26E-07  \\
 			13 & 0 & 0.0168123 & 0 & 1.00E-06 & 0 & 8.21E-07  \\
 			14 & 0 & 0.0167919 & 0 & 9.98E-07 & 0 & 8.19E-07  \\
 			15 & 0 & 0.0167817 & 0 & 9.97E-07 & 0 & 8.18E-07  \\
 			16 & 0 & 0.0167766 & 0 & 9.96E-07 & 0 & 8.17E-07  \\
 			17 & 0 & 0.0167741 & 0 & 9.96E-07 & 0 & 8.17E-07  \\
 			18 & 0 & 0.0167728 & 0 & 9.95E-07 & 0 & 8.17E-07  \\
 			19 & 0 & 0.0167722 & 0 & 9.95E-07 & 0 & 8.17E-07  \\
 			20 & \underline{4} & 0.0167719 & 0 & 9.95E-07 & 0 & 8.17E-07  \\
 			\hline
 		\end{tabular}
 	\end{center}
 	\label{tab:ppctriotest}
 \end{table}
 
 \clearpage
 We conclude from this that the test dataset is too small and unvaried enough in genotypes to allow the random generation of a good dataset of mother-father-child trios. We therefore do not pursue clustering algorithms using this simulated dataset of family trios in this paper.

\section{Clustering simulations}

\subsection{Single group simulations}

Foe each of the 100 instances of the four two-person relationships, 5 admixtures were simulated in which each contributor contributed five of the simulated \scepgs. The admixtures were the clustered using the partition search, using a population sub-structure parameter of $\theta=0.02$. The results are summarized in \tabref{tab:simsum}. In only two of the 2000 simulated admixtures did the clustering algorithm \textit{not} have the true partition as the highest likelihood partition. 

We see from \tabref{tab:simsum} that in the majority of cases the estimated distance was correctly found, and if not, within an error of $\pm 1$ in most other cases.

\begin{table}[htbp]
	\caption{Summary of estimated maximized distance counts for the total of 500 simulations on each of the two-person relationships, in which the highest likelihood partition found by the clustering algorithm was a correct partition. There were just two cases for which this was not the case, both in the Full-sibs simulations.\label{tab:simsum}}
	\begin{center}
		\begin{tabular}{c|cccc}\hline
			$\hat{k}$ &Full-Sibs & Parent-Child & Half-Sibs & Grandparent-grandchild  \\ \hline 
			1 & 19 & \underline{434} & 29 & 22  \\ 
			1.5 & \underline{455} & 65 & 142 & 108  \\ 
			2 & 24 & 1 & \underline{248} & \underline{237}  \\ 
			3 & - & - & 57 & 77  \\ 
			4 & - & - & 15 & 34  \\ 
			5 & - & - & 1 & 17  \\ 
			6 & - & - & 1 & 3  \\ 
			10 & - & - & 7 & 2  \\ 
			total & 498 & 500 & 500 & 500  \\ 
			\hline
		\end{tabular}
	\end{center}
	\label{tab:simre2}
\end{table}

The two Full-Sibs simulations, in which the highest likelihood partition was not the true partition, are interesting anomalies.
In the searches, at each stage only partitions with likelihood values within  10 on the log-10 scale, of the highest likelihood partition were retained. They both occurred for the two different simulated admixtures for the same pair of simulated persons: the simulated genotypes are shown in \tabref{tab:simgt}. We see that they have at least one allele in common on every locus, and on 14 of the loci they have the same genotype.

\begin{table}[htbp]
	\caption{Genotypes of the two simulated persons for the two partition searches which did not find the highest likelihood partition to be the true partition.}
	\begin{center}
		\begin{tabular}{lcc|lcc}\hline
			RP\_1 & allele1 & allele2 & RP\_2 & allele1 & allele2  \\ \hline 
CSF1PO & 12 & 13 & CSF1PO & 12 & 13  \\ 
D10S1248 & 13 & 14 & D10S1248 & 14 & 16  \\ 
D12S391 & 17 & 21 & D12S391 & 17 & 21  \\ 
D13S317 & 11 & 12 & D13S317 & 11 & 12  \\ 
D16S539 & 12 & 9 & D16S539 & 12 & 9  \\ 
D18S51 & 14 & 18 & D18S51 & 14 & 14  \\ 
D19S433 & 13 & 14 & D19S433 & 13 & 14  \\ 
D1S1656 & 12 & 17.3 & D1S1656 & 12 & 17.3  \\ 
D21S11 & 30 & 31.2 & D21S11 & 30 & 31.2  \\ 
D22S1045 & 15 & 15 & D22S1045 & 15 & 15  \\ 
D2S1338 & 17 & 23 & D2S1338 & 17 & 17  \\ 
D2S441 & 11 & 14 & D2S441 & 11 & 14  \\ 
D3S1358 & 15 & 15 & D3S1358 & 15 & 15  \\ 
D5S818 & 10 & 11 & D5S818 & 11 & 11  \\ 
D7S820 & 11 & 8 & D7S820 & 11 & 9  \\ 
D8S1179 & 13 & 14 & D8S1179 & 13 & 14  \\ 
FGA & 21 & 22 & FGA & 21 & 22  \\ 
SE33 & 20 & 28.2 & SE33 & 20 & 28.2  \\ 
TH01 & 6 & 9 & TH01 & 6 & 9  \\ 
TPOX & 11 & 8 & TPOX & 11 & 8  \\ 
vWA & 16 & 17 & vWA & 15 & 17  \\ 
			\hline
		\end{tabular}
	\end{center}
	\label{tab:simgt}
\end{table}

Also, interestingly, for both of these two simulated admixtures, the highest likelihood partition had a single cluster, having base-10 log-likelihoods of -1523.07 and -1498.77 respectively. Given their genotypes, perhaps  this is not so surprising. Looking beyond the highest likelihood partition for each admixture, the log-likelihoods for the top 11 partitions found for each admixture are shown in \tabref{tab:toplls}, from which we see that they are closely bunched.

\begin{table}[htbp]
	\caption{Log-likelihoods (base 10) of the top 11 partitions found for both admixture, for which the highest likelihood partition was a single cluster.}
	\begin{center}
		\begin{tabular}{c|c}\hline
			Admixture 1 & Admixture 2  \\ \hline 
			-1523.07	&	-1498.78	\\
			-1523.76	&	-1499.43	\\
			-1525.61	&	-1500.81	\\
			-1527.05	&	-1501.57	\\
			-1527.98	&	-1501.81	\\
			-1528.04	&	-1502.10	\\
			-1528.13	&	-1502.10	\\
			-1528.42	&	-1502.85	\\
			-1528.82	&	-1503.10	\\
			-1528.87	&	-1503.39	\\
			-1529.37	&	-1503.87	\\
%
			\hline
		\end{tabular}
	\end{center}
	\label{tab:toplls}
\end{table}

For both admixtures, in all of the partitions found except the highest likelihoods ones there were two clusters. 
For one admixture the true partition was the second highest likelihood partition. For the other admixture,  the true partition was not among the 22 partitions retained; however in all the cases in one of the pair of clusters person RP\_2 was the only contributor, with the other cluster having cells from both contributors.

However, when taking into account the change of likelihood when incorporating the estimated relationship distance between the two clusters in each partition, we see that familial relationship is now favoured for both admixtures. In \tabref{tab:admix1} shows the adjustments for the first admixture, for the ten highest likelihood partitions, and similarly \tabref{tab:admix2} shows those for the second admixture. The first line of each tables shows the log-likelihood for the highest scoring partition consisting in both cases of a single cluster, which does not have an adjustment. In both cases we case see that inclusion of possible relatedness is highly favoured for many of the other partitions over the single cluster partition (that does not get a relatedness adjustment).

\begin{table}[htbp]
	\caption{Admixture 1 partition log10-likelihoods with  computed adjustments for relatedness. The largest adjusted likelihood is on line 2. }
	\begin{center}
		\begin{tabular}{ccc}\hline 
			Original LL & adjustment & total adjusted LL \\ \hline
			-1523.07	&	0	&	-1523.07	  \\ \hline
			-1523.76	&	8.75	&	-1515.01	\\
			-1525.61	&	8.07	&	-1517.54	\\
			-1527.05	&	9.34	&	-1517.71	\\
			-1527.98	&	7.19	&	-1520.79	\\
			-1528.04	&	8.74	&	-1519.30	\\
			-1528.13	&	5.48	&	-1522.65	\\
			-1528.42	&	7.85	&	-1520.57	\\
			-1528.82	&	8.68	&	-1520.14	\\
			-1528.87	&	4.47	&	-1524.40	\\
			-1529.37	&	8.33	&	-1521.04	\\
%
%
			\hline
		\end{tabular}
	\end{center}
	\label{tab:admix1}
\end{table}

\begin{table}[htbp]
	\caption{Admixture 2 partition log10-likelihoods with  computed adjustments for relatedness. The largest adjusted likelihood is on line 2.}
	\begin{center}
		\begin{tabular}{ccc}\hline
				Original LL & adjustment & total adjusted LL \\ \hline
				-1498.78	&	0	&	-1498.78	\\	\hline
				-1499.43	&	10.25	&	-1489.18	\\	
				-1500.81	&	5.14	&	-1495.67	\\	
				-1501.57	&	10.23	&	-1491.34	\\	
				-1501.81	&	10.24	&	-1491.57	\\	
				-1502.10	&	9.26	&	-1492.84	\\	
				-1502.10	&	9.48	&	-1492.62	\\	
				-1502.85	&	7.53	&	-1495.32	\\	
				-1503.10	&	8.59	&	-1494.51	\\	
				-1503.39	&	9.59	&	-1493.80	\\	
				-1503.87	&	9.23	&	-1494.64	\\	
%
%
%
			\hline
		\end{tabular}
	\end{center}
	\label{tab:admix2}
\end{table}

\subsection{A three group simulation}

In the simulations described above, all simulated mixtures were obtained using two contributors who were closely related. In the simulation of this section, 100 sets of profiles of six simulated people, labelled $RP1, RP2, \dots, RP6$ and \scepgs simulated for each person based on their genotypes were formed, where there were the following three pairwise relationships simulated:

\begin{enumerate}
	\item $RP1$ and $RP2$: Full siblings ($k=1.5)$;
	\item $RP3$ and $RP4$: Parent-child ($k=1)$;
	\item $RP5$ and $RP6$: Half-sibling (or grandparent-grandchild) ($k=2$);
\end{enumerate}

By the method of their construction, apart from these relationships all other pairs of simulated people are unrelated.

For each os the 100 sets of genotypes and simulated \scepgs, an admixtures was made by sampling without replacement 5 simulated \scepgs from each person. The partition algorithm was then run, in which only those \scepgs of sufficient quality, as recorded by the total number of larger peaks in the \scepg, were retained. It was found that in all simulations that the highest likelihood partition  was the true partition \textit{for the \scepgs retained}: it was found in some simulations that a person contributed no \scepgs of sufficient quality to be retained for the partition search.

For each top partition, the estimated IBD distances of each pair of clusters was found. It was known which simulated person was associated with each cluster, so that for pairs of clusters associated with two simulated persons, the counts of the number of estimated distances from $k=1$ to 10 were stored; when the pair of clusters were not from pairs of related people, these counts were stored as an unrelated pair $U,U$ pair.
The table of counts are shown in \tabref{tab:sixcounts}. The final line shows what the count total would have been if all six persons contributed cells to the final highest likelihood partition.

\begin{table}[ht]
	\caption{Pairwise distance ($k$)counts for the 100 simulations of
		admixtures of six persons, with those for the simulated pairwise relationships, and all other groups in the final column for two unrelated persons. Distances were estimated up to a value of 10, which is taken to represent unrelated in this simulation. Underlined counts are for the correct distances.\label{tab:sixcounts}}
		\begin{center}
\begin{tabular}{l|cccc}
	$\hat{k}$  & RP1,RP2 & RP3,RP4 & RP5,RP6 & U,U \\ \hline
	1 & 11 & \underline{92} & 7 & 1\\
	1.5 & \underline{64} & 1 & 2 & 4\\
	2 & 15 & 0 & \underline{38} & 92\\
	3 & 2 & 0 & 24 & 161\\
	4 & 0 & 0 & 9 & 121\\
	5 & 0 & 0 & 2 & 78\\
	6 & 0 & 0 & 1 & 45\\
	7 & 0 & 0 & 1 & 26\\
	8 & 0 & 0 & 0 & 8\\
	9 & 0 & 0 & 0 & 8\\
	10 & 1 & 0 & 6 & \underline{549}\\ \hline
	Total &93 & 93 & 90 & 1093\\
	Out of &100 & 100 &100 &  1200 \\ \hline
	
\end{tabular}
\end{center}
\end{table}

 We see from the table that, in the main, the correct distance is found exactly or closely. The $U,U$ column suggests that the simulation of the persons, from the limited set of profiles in the real data, may be skewing the results. Otherwise, this simulation appears to show that small groups of closely related individuals may be identifiable using the method proposed in this paper.

\section{Clustering simulations}

\subsection{Single group simulations}

Foe each of the 100 instances of the four two-person relationships, 5 admixtures were simulated in which each contributor contributed five of the simulated \scepgs. The admixtures were the clustered using the partition search, using a population sub-structure parameter of $\theta=0.02$. The results are summarized in \tabref{tab:simsum}. In only two of the 2000 simulated admixtures did the clustering algorithm \textit{not} have the true partition as the highest likelihood partition. 

We see from \tabref{tab:simsum} that in the majority of cases the estimated distance was correctly found, and if not, within an error of $\pm 1$ in most other cases.

\begin{table}[htbp]
	\caption{Summary of estimated maximized distance counts for the total of 500 simulations on each of the two-person relationships, in which the highest likelihood partition found by the clustering algorithm was a correct partition. There were just two cases for which this was not the case, both in the Full-sibs simulations.\label{tab:simsum}}
	\begin{center}
		\begin{tabular}{c|cccc}\hline
			$\hat{k}$ &Full-Sibs & Parent-Child & Half-Sibs & Grandparent-grandchild  \\ \hline 
			1 & 19 & \underline{434} & 29 & 22  \\ 
			1.5 & \underline{455} & 65 & 142 & 108  \\ 
			2 & 24 & 1 & \underline{248} & \underline{237}  \\ 
			3 & - & - & 57 & 77  \\ 
			4 & - & - & 15 & 34  \\ 
			5 & - & - & 1 & 17  \\ 
			6 & - & - & 1 & 3  \\ 
			10 & - & - & 7 & 2  \\ 
			total & 498 & 500 & 500 & 500  \\ 
			\hline
		\end{tabular}
	\end{center}
	\label{tab:simre2}
\end{table}

The two Full-Sibs simulations, in which the highest likelihood partition was not the true partition, are interesting anomalies.
In the searches, at each stage only partitions with likelihood values within  10 on the log-10 scale, of the highest likelihood partition were retained. They both occurred for the two different simulated admixtures for the same pair of simulated persons: the simulated genotypes are shown in \tabref{tab:simgt}. We see that they have at least one allele in common on every locus, and on 14 of the loci they have the same genotype.

\begin{table}[htbp]
	\caption{Genotypes of the two simulated persons for the two partition searches which did not find the highest likelihood partition to be the true partition.}
	\begin{center}
		\begin{tabular}{lcc|lcc}\hline
			RP\_1 & allele1 & allele2 & RP\_2 & allele1 & allele2  \\ \hline 
CSF1PO & 12 & 13 & CSF1PO & 12 & 13  \\ 
D10S1248 & 13 & 14 & D10S1248 & 14 & 16  \\ 
D12S391 & 17 & 21 & D12S391 & 17 & 21  \\ 
D13S317 & 11 & 12 & D13S317 & 11 & 12  \\ 
D16S539 & 12 & 9 & D16S539 & 12 & 9  \\ 
D18S51 & 14 & 18 & D18S51 & 14 & 14  \\ 
D19S433 & 13 & 14 & D19S433 & 13 & 14  \\ 
D1S1656 & 12 & 17.3 & D1S1656 & 12 & 17.3  \\ 
D21S11 & 30 & 31.2 & D21S11 & 30 & 31.2  \\ 
D22S1045 & 15 & 15 & D22S1045 & 15 & 15  \\ 
D2S1338 & 17 & 23 & D2S1338 & 17 & 17  \\ 
D2S441 & 11 & 14 & D2S441 & 11 & 14  \\ 
D3S1358 & 15 & 15 & D3S1358 & 15 & 15  \\ 
D5S818 & 10 & 11 & D5S818 & 11 & 11  \\ 
D7S820 & 11 & 8 & D7S820 & 11 & 9  \\ 
D8S1179 & 13 & 14 & D8S1179 & 13 & 14  \\ 
FGA & 21 & 22 & FGA & 21 & 22  \\ 
SE33 & 20 & 28.2 & SE33 & 20 & 28.2  \\ 
TH01 & 6 & 9 & TH01 & 6 & 9  \\ 
TPOX & 11 & 8 & TPOX & 11 & 8  \\ 
vWA & 16 & 17 & vWA & 15 & 17  \\ 
			\hline
		\end{tabular}
	\end{center}
	\label{tab:simgt}
\end{table}

Also, interestingly, for both of these two simulated admixtures, the highest likelihood partition had a single cluster, having base-10 log-likelihoods of -1523.07 and -1498.77 respectively. Given their genotypes, perhaps  this is not so surprising. Looking beyond the highest likelihood partition for each admixture, the log-likelihoods for the top 11 partitions found for each admixture are shown in \tabref{tab:toplls}, from which we see that they are closely bunched.

\begin{table}[htbp]
	\caption{Log-likelihoods (base 10) of the top 11 partitions found for both admixture, for which the highest likelihood partition was a single cluster.}
	\begin{center}
		\begin{tabular}{c|c}\hline
			Admixture 1 & Admixture 2  \\ \hline 
			-1523.07	&	-1498.78	\\
			-1523.76	&	-1499.43	\\
			-1525.61	&	-1500.81	\\
			-1527.05	&	-1501.57	\\
			-1527.98	&	-1501.81	\\
			-1528.04	&	-1502.10	\\
			-1528.13	&	-1502.10	\\
			-1528.42	&	-1502.85	\\
			-1528.82	&	-1503.10	\\
			-1528.87	&	-1503.39	\\
			-1529.37	&	-1503.87	\\
%
			\hline
		\end{tabular}
	\end{center}
	\label{tab:toplls}
\end{table}

For both admixtures, in all of the partitions found except the highest likelihoods ones there were two clusters. 
For one admixture the true partition was the second highest likelihood partition. For the other admixture,  the true partition was not among the 22 partitions retained; however in all the cases in one of the pair of clusters person RP\_2 was the only contributor, with the other cluster having cells from both contributors.

However, when taking into account the change of likelihood when incorporating the estimated relationship distance between the two clusters in each partition, we see that familial relationship is now favoured for both admixtures. In \tabref{tab:admix1} shows the adjustments for the first admixture, for the ten highest likelihood partitions, and similarly \tabref{tab:admix2} shows those for the second admixture. The first line of each tables shows the log-likelihood for the highest scoring partition consisting in both cases of a single cluster, which does not have an adjustment. In both cases we case see that inclusion of possible relatedness is highly favoured for many of the other partitions over the single cluster partition (that does not get a relatedness adjustment).

\begin{table}[htbp]
	\caption{Admixture 1 partition log10-likelihoods with  computed adjustments for relatedness. The largest adjusted likelihood is on line 2. }
	\begin{center}
		\begin{tabular}{ccc}\hline 
			Original LL & adjustment & total adjusted LL \\ \hline
			-1523.07	&	0	&	-1523.07	  \\ \hline
			-1523.76	&	8.75	&	-1515.01	\\
			-1525.61	&	8.07	&	-1517.54	\\
			-1527.05	&	9.34	&	-1517.71	\\
			-1527.98	&	7.19	&	-1520.79	\\
			-1528.04	&	8.74	&	-1519.30	\\
			-1528.13	&	5.48	&	-1522.65	\\
			-1528.42	&	7.85	&	-1520.57	\\
			-1528.82	&	8.68	&	-1520.14	\\
			-1528.87	&	4.47	&	-1524.40	\\
			-1529.37	&	8.33	&	-1521.04	\\
%
%
			\hline
		\end{tabular}
	\end{center}
	\label{tab:admix1}
\end{table}

\begin{table}[htbp]
	\caption{Admixture 2 partition log10-likelihoods with  computed adjustments for relatedness. The largest adjusted likelihood is on line 2.}
	\begin{center}
		\begin{tabular}{ccc}\hline
				Original LL & adjustment & total adjusted LL \\ \hline
				-1498.78	&	0	&	-1498.78	\\	\hline
				-1499.43	&	10.25	&	-1489.18	\\	
				-1500.81	&	5.14	&	-1495.67	\\	
				-1501.57	&	10.23	&	-1491.34	\\	
				-1501.81	&	10.24	&	-1491.57	\\	
				-1502.10	&	9.26	&	-1492.84	\\	
				-1502.10	&	9.48	&	-1492.62	\\	
				-1502.85	&	7.53	&	-1495.32	\\	
				-1503.10	&	8.59	&	-1494.51	\\	
				-1503.39	&	9.59	&	-1493.80	\\	
				-1503.87	&	9.23	&	-1494.64	\\	
%
%
%
			\hline
		\end{tabular}
	\end{center}
	\label{tab:admix2}
\end{table}

\subsection{A three group simulation}

In the simulations described above, all simulated mixtures were obtained using two contributors who were closely related. In the simulation of this section, 100 sets of profiles of six simulated people, labelled $RP1, RP2, \dots, RP6$ and \scepgs simulated for each person based on their genotypes were formed, where there were the following three pairwise relationships simulated:

\begin{enumerate}
	\item $RP1$ and $RP2$: Full siblings ($k=1.5)$;
	\item $RP3$ and $RP4$: Parent-child ($k=1)$;
	\item $RP5$ and $RP6$: Half-sibling (or grandparent-grandchild) ($k=2$);
\end{enumerate}

By the method of their construction, apart from these relationships all other pairs of simulated people are unrelated.

For each os the 100 sets of genotypes and simulated \scepgs, an admixtures was made by sampling without replacement 5 simulated \scepgs from each person. The partition algorithm was then run, in which only those \scepgs of sufficient quality, as recorded by the total number of larger peaks in the \scepg, were retained. It was found that in all simulations that the highest likelihood partition  was the true partition \textit{for the \scepgs retained}: it was found in some simulations that a person contributed no \scepgs of sufficient quality to be retained for the partition search.

For each top partition, the estimated IBD distances of each pair of clusters was found. It was known which simulated person was associated with each cluster, so that for pairs of clusters associated with two simulated persons, the counts of the number of estimated distances from $k=1$ to 10 were stored; when the pair of clusters were not from pairs of related people, these counts were stored as an unrelated pair $U,U$ pair.
The table of counts are shown in \tabref{tab:sixcounts}. The final line shows what the count total would have been if all six persons contributed cells to the final highest likelihood partition.

\begin{table}[ht]
	\caption{Pairwise distance ($k$)counts for the 100 simulations of
		admixtures of six persons, with those for the simulated pairwise relationships, and all other groups in the final column for two unrelated persons. Distances were estimated up to a value of 10, which is taken to represent unrelated in this simulation. Underlined counts are for the correct distances.\label{tab:sixcounts}}
		\begin{center}
\begin{tabular}{l|cccc}
	$\hat{k}$  & RP1,RP2 & RP3,RP4 & RP5,RP6 & U,U \\ \hline
	1 & 11 & \underline{92} & 7 & 1\\
	1.5 & \underline{64} & 1 & 2 & 4\\
	2 & 15 & 0 & \underline{38} & 92\\
	3 & 2 & 0 & 24 & 161\\
	4 & 0 & 0 & 9 & 121\\
	5 & 0 & 0 & 2 & 78\\
	6 & 0 & 0 & 1 & 45\\
	7 & 0 & 0 & 1 & 26\\
	8 & 0 & 0 & 0 & 8\\
	9 & 0 & 0 & 0 & 8\\
	10 & 1 & 0 & 6 & \underline{549}\\ \hline
	Total &93 & 93 & 90 & 1093\\
	Out of &100 & 100 &100 &  1200 \\ \hline
	
\end{tabular}
\end{center}
\end{table}

 We see from the table that, in the main, the correct distance is found exactly or closely. The $U,U$ column suggests that the simulation of the persons, from the limited set of profiles in the real data, may be skewing the results. Otherwise, this simulation appears to show that small groups of closely related individuals may be identifiable using the method proposed in this paper.

 \section{Summary}

We have proposed combining the clustering algorithm of \cite{cowell2025clustering} for analysing \scepgs from autosomal STR DNA mixtures with the pedigree search method of \cite{cowell2003clustering} to look for related individuals in mixtures using single cell analysis. In the absence of an existing dataset of DNA from related individuals, it was proposed to use artificial person profiles simulated from a dataset of \scepgs\ from real individuals, and to simulate \scepgs\ from the same dataset by treating the peak height information on the \scepg loci as interchangeable within each locus, and sampling the without replacement to generate simulated \scepgs. It was shown that the combined methods produced good results for closely related pairs of individuals using the simulated data, indicating it could be a promising route to dealing with related individuals when analysing DNA mixtures using single cell data. It remains to be seen if the results of this paper will be confirmed on real data  based on DNA samples from real related individuals, thus avoiding the limitations discussed earlier with the simulated people and \scepgs used in this paper, which as previously mentioned may have biased the results.

\section*{Acknowledgments}

The  author would like to thank Catherine Grgicak and Desmond Lun for   confidential access to 
the single cell data used in this paper, and additionally to these and  also Klaas Slooten for feedback on the approach of this paper.

\appendix

\section{IBD distance with population sub-structure}
\label{sec:app1}

Here we extend the formula in \cite{cowell2003clustering} from Hardy-Weinberg equilibrium to include population sub-structure.
We define the following two functions

\begin{align*}
	F(n,p_a,\theta) &= \prod_{i=1}^n [ (1-\theta)p_a + (i-1)\theta ], & F(n,p,0) &= p^n,\\
	G(n,\theta) &= \prod_{i=1}^n [ (1-\theta) + (i-1)\theta ], &
	G(n,0) &=1,\\
\end{align*}

where $\theta$ is the population sub-structure parameter.

In terms of these, Table~1 of \cite{cowell2003clustering} is modified
to the following table, which defines a number of $H$ functions.

\begin{table}[ht]
	\centering
\resizebox{\textwidth}{!}{
		\begin{tabular}{|l|c|c|c|} \hline
			Cases: & None IBD      & One pair IBD      & Two pairs IBD \\ \hline
			(1) aa, aa  \phantom{\Large $\int_x^x$}
			& $H_{1,0}=\frac{F(4,p_a,\theta)}{G(4,\theta)}$         & $H_{1,1}=\frac{F(3,p_a,\theta)}{G(3,\theta)}$             &  $H_{1,2}=\frac{F(2,p_a,\theta)}{G(2,\theta)}$         \\ \hline
			(2) aa, ab  \phantom{\Large $\int_x^x$}
			& $H_{2,0}=4\frac{F(3,p_a,\theta)F(1,p_b,\theta)}{G(4,\theta)}$     & $H_{2,1}=2\frac{F(2,p_a,\theta)F(1,p_b,\theta)}{G(3,\theta)}$         & 0             \\ \hline
			(3) ab, ab  \phantom{\Large $\int_x^x$}
			& $H_{3,0}=4\frac{F(2,p_a,\theta)F(2,p_b,\theta)}{G(4,\theta)}$   &
			$H_{3,1}=\frac{F(2,p_a,\theta)F(1,p_b,\theta) + F(1,p_a,\theta)F(2,p_b,\theta)	}{G(3,\theta)}$   &
			$H_{3,2}=2\frac{F(1,p_a,\theta)F(1,p_b,\theta)}{G(2,\theta)}$
			\\ \hline
			(4) ab, ac  \phantom{\Large $\int_x^x$}
			& 
			$H_{4,0}=8\frac{F(2,p_a,\theta)F(1,p_b,\theta)F(1,p_c,\theta)}{G(4,\theta)}$ 
			& $H_{4,1}=2\frac{F(1,p_a,\theta)F(1,p_b,\theta)F(1,p_c,\theta)}{G(3,\theta)}$         & 0             \\ \hline
			(5) no alleles in common  \phantom{\Large $\int_x^x$}
			& (depends) & 0       & 0                \\ \hline       
		\end{tabular}
}
\end{table}

Following the same line of arguments in that paper, we arrive at the following that generalizes Table~3 of \cite{cowell2003clustering} for likelihood ratios of distance  versus infinite distance for various combinations of genotype configurations, in the absence of inbreeding:

\begin{table}[htbp]
	\begin{center}
		\begin{tabular}{|l|c|c|}  \hline
			Cases: & Distance $i\in{\mathbb N}$ & Distance 1.5 \\ \hline
			(1) aa, aa  \phantom{\Large $\int_x^x$}
			& $(\frac12)^{i-1}\frac{H_{1,1}}{H_{1,0}}+1-(\frac12)^{i-1}$ & $\frac14\frac{H_{1,2}}{H_{1,0}}+\frac12\frac{H_{1,1}}{H_{1,0}}+\frac14$ \\ \hline
			(2) aa, ab  \phantom{\Large $\int_x^x$}
			& $(\frac12)^{i-1}\frac{H_{2,1}}{H_{2,0}}+1-(\frac12)^{i-1}$ & $\frac12\frac{H_{2,1}}{H_{2,0}}+\frac14$ \\ \hline
			(3) ab, ab  \phantom{\Large $\int_x^x$}
			& $(\frac12)^{i-1}\frac{H_{3,1}}{H_{3,0}}+1-(\frac12)^{i-1}$ & $\frac14\frac{H_{3,2}}{H_{3,0}}+\frac12\frac{H_{3,1}}{H_{3,0}}+\frac14$ \\ \hline
			(4) ab, ac  \phantom{\Large $\int_x^x$}
			& $(\frac12)^{i-1}\frac{H_{4,1}}{H_{4,0}}+1-(\frac12)^{i-1}$ & $\frac12\frac{H_{4,1}}{H_{4,0}}+\frac14$ \\ \hline
			(5) no alleles in common \phantom{\Large $\int_x^x$}
			& $1 - (\frac12)^{i-1}$ & $\frac14$   \\ \hline
		\end{tabular}
	\end{center}
\end{table}


\begin{thebibliography}{14}
\providecommand{\natexlab}[1]{#1}
\providecommand{\url}[1]{\texttt{#1}}
\expandafter\ifx\csname urlstyle\endcsname\relax
  \providecommand{\doi}[1]{doi: #1}\else
  \providecommand{\doi}{doi: \begingroup \urlstyle{rm}\Url}\fi

\bibitem[Ballantyne et~al.(2026)Ballantyne, Grgicak, and
  Marciano]{ballantyne2026next}
Jack Ballantyne, Catherine~M Grgicak, and Michael~A Marciano.
\newblock The next frontier in forensic genetics: Single-cell genetic analysis
  and its expanding role.
\newblock \emph{Forensic Science International: Genetics}, page 103568, 2026.

\bibitem[Bhembe et~al.(2025)Bhembe, Lun, Duffy, and Grgicak]{qhawe2025}
Qhawe~A. Bhembe, Desmond~S. Lun, Ken~R. Duffy, and Catherine~M. Grgicak.
\newblock Salience, legitimacy, and credibility of two end-to end forensic
  single cell probabilistic systems.
\newblock \emph{Forensic Science International:Genetics}, 2025.
\newblock \doi{https://doi.org/10.1016/j.fsigen.2025.103369}.

\bibitem[Bleka et~al.(2025)Bleka, Vigeland, and Gill]{bleka2025efmex}
{\O}yvind Bleka, Magnus~Dehli Vigeland, and Peter Gill.
\newblock {EFMex}: Using {EuroForMix} to evaluate {DNA} mixtures with multiple
  persons of interest.
\newblock \emph{Forensic Science International: Genetics}, page 103365, 2025.

\bibitem[Butler et~al.(2003)Butler, Schoske, Vallone, Redman, and
  Kline]{butler:etal:03}
J.~M. Butler, R.~Schoske, P.~M. Vallone, J.~W. Redman, and M.~C. Kline.
\newblock Allele frequencies for 15 autosomal loci on {U.S.} {C}aucasian,
  {A}frican {A}merican, and {H}ispanic populations.
\newblock \emph{Journal of Forensic Science}, 48\penalty0 (4):\penalty0
  908--911, 2003.
\newblock Available online at www.astm.org.

\bibitem[Cowell(2025)]{cowell2025clustering}
Robert~G Cowell.
\newblock A clustering algorithm for the single cell analysis of mixtures.
\newblock \emph{arXiv preprint arXiv:2510.10614}, 2025.

\bibitem[Cowell and Mostad(2003)]{cowell2003clustering}
Robert~G Cowell and Petter Mostad.
\newblock A clustering algorithm using {DNA} marker information for
  sub-pedigree reconstruction.
\newblock \emph{Journal of forensic sciences}, 48\penalty0 (6):\penalty0
  1239--1248, 2003.

\bibitem[D{\o}rum et~al.(2017)D{\o}rum, Kaur, and Gysi]{Dorum2017}
G.~D{\o}rum, N.~Kaur, and M.~Gysi.
\newblock Pedigree-based relationship inference from complex {DNA} mixtures.
\newblock \emph{International Journal of Legal Medicine}, 131:\penalty0
  629--641, 2017.
\newblock \doi{10.1007/s00414-016-1526-x}.

\bibitem[Duffy et~al.(2023)Duffy, Lun, Mulcahy, O’Donnell, Sheth, and
  Grgicak]{duffy2023evidentiary}
Ken~R Duffy, Desmond~S Lun, Madison~M Mulcahy, Leah O’Donnell, Nidhi Sheth,
  and Catherine~M Grgicak.
\newblock Evidentiary evaluation of single cells renders highly informative
  forensic comparisons across multifarious admixtures.
\newblock \emph{Forensic Science International: Genetics}, 64:\penalty0 102852,
  2023.

\bibitem[Gill et~al.(1994)Gill, Ivanov, Kimpton, Piercy, Benson, Tully, Evett,
  Hagelberg, and Sullivan]{gill1994identification}
Peter Gill, Pavel~L Ivanov, Colin Kimpton, Romelle Piercy, Nicola Benson,
  Gillian Tully, Ian Evett, Erika Hagelberg, and Kevin Sullivan.
\newblock Identification of the remains of the {R}omanov family by {DNA}
  analysis.
\newblock \emph{Nature genetics}, 6\penalty0 (2):\penalty0 130--135, 1994.

\bibitem[Green(2020)]{Green2020KinMix}
Peter~J. Green.
\newblock {KinMix}: {DNA} mixture analysis with related contributors.
\newblock \url{https://petergreenweb.wordpress.com/kinmix2-0}, 2020.
\newblock Software package.

\bibitem[Green and Mortera(2021)]{GreenMortera2021}
Peter~J. Green and Julia Mortera.
\newblock Inference about complex relationships using peak height data from
  {DNA} mixtures.
\newblock \emph{Journal of the Royal Statistical Society: Series C (Applied
  Statistics)}, 70\penalty0 (4):\penalty0 1049--1082, 2021.
\newblock \doi{10.1111/rssc.12498}.

\bibitem[Hernandis et~al.(2019)Hernandis, D{\o}rum, and
  Egeland]{Hernandis2019Relmix}
E.~Hernandis, G.~D{\o}rum, and T.~Egeland.
\newblock Relmix: an open source software for {DNA} mixtures with related
  contributors.
\newblock In \emph{Forensic Science International: Genetics Supplement Series},
  volume~7, pages 221--223, 2019.
\newblock \doi{10.1016/j.fsigss.2019.09.085}.

\bibitem[Taylor et~al.(2013)Taylor, Bright, and Buckleton]{Taylor2013}
D.~Taylor, J.-A. Bright, and J.~Buckleton.
\newblock The interpretation of single source and mixed {DNA} profiles.
\newblock \emph{Forensic Science International: Genetics}, 7:\penalty0
  516--528, 2013.
\newblock \doi{10.1016/j.fsigen.2013.05.011}.

\bibitem[Taylor et~al.(2014)Taylor, Bright, and Buckleton]{Taylor2014}
D.~Taylor, J.-A. Bright, and J.~Buckleton.
\newblock Considering relatives when assessing the evidential strength of mixed
  {DNA} profiles.
\newblock \emph{Forensic Science International: Genetics}, 13:\penalty0
  259--263, 2014.
\newblock \doi{10.1016/j.fsigen.2014.08.015}.

\end{thebibliography}
\end{document}